\documentstyle[aps,preprint,epsfig]{revtex}

\draft

\begin{document}

\title{\bf Towards a tomographic picture of a Bose-Einstein condensate}
\author{Stefano~Mancini} 
\address{Istituto Nazionale per la Fisica della Materia,
         \\ Dipartimento di Fisica, Universit\`a di Milano,
         Via Celoria 16, I-20133 Milano, Italy}
\author{Mauro~Fortunato}  
\address{Istituto Nazionale per la Fisica della Materia,
         \\ Dipartimento di Matematica e Fisica, Universit\'a di Camerino, 
         I-62032 Camerino, Italy}
\author{Paolo~Tombesi}
\address{Istituto Nazionale per la Fisica della Materia,
         \\ Dipartimento di Matematica e Fisica, Universit\'a di Camerino, 
         I-62032 Camerino, Italy}
\author{Giacomo Mauro D'Ariano}
\address{Istituto Nazionale per la Fisica della Materia,
         \\ Dipartimento di Fisica ``A. Volta", Universit\`a di Pavia,
         I-27100 Pavia, Italy}
\date{Received: \today}
\maketitle
 
\begin{abstract}
We investigate by means of numerical simulations 
the possibilities of tomographic techniques
applied to a Bose-Einstein condensate in order 
to reconstruct its ground state.
Essentially, two scenarios are considered for which 
the density matrix elements can be retrieved from 
atom counting probabilities.
The methods presented here allow to distinguish among
various possible quantum states.
\end{abstract}

\pacs{OCIS key terms: Quantum Optics 270.0270, Atomic and molecular 
physics 020.0020 \\ PACS number(s): 03.65.Bz, 03.75.Fi, 32.80.-t}

\section{Introduction}

Before the birth of quantum mechanics, optics and mechanics have long 
developed on parallel tracks, as light and massive entities were 
considered as waves and particles, respectively. At the beginning of 
the 20th century, with the introduction of quantum mechanics, waves 
and particles started to play an interchangeable role, with the 
concepts of photons and of De Broglie wavelength. This gave rise to 
the birth of quantum optics and atom optics~\cite{miwa}.
However, while an optical single-mode system has already been 
available since long ago, the same cannot be said for matter waves.
In fact, in the field of atom optics, only recently breakthroughs in the
evaporative cooling of dilute alkali gases have
allowed the generation of Bose-Einstein condensates (BEC) \cite{exp}.
The BEC is a macroscopic occupation of the ground state of the gas and is one
important paradigm of quantum statistical mechanics.

In recent theoretical and experimental investigations of BEC, one of the most
important and urgent issues is the determination of the actual quantum state of 
the condensate.  
In fact, as in optics, the presence of many particles in a single mode makes it 
possible to inquire about the multiparticle quantum state of the mode.
At first thought, a number state might seem a natural description of a
condensate mode, but the actual state may well depend on the details of
preparation. For example the demonstration of first-order interference
\cite{Ketterle} and observations of normalized, spatial correlation functions
\cite{corr} near unity
suggest coherent states, while the presence of collisions between atoms may lead to the formation of
squeezed states
\cite{squee}. Also the internal states of the condensate atoms allow precise manipulation of the BEC
state by interaction with light \cite{Zeng}.

Few experimental methods of obtaining partial information about the state have been suggested
\cite{stateinfo}. 
On the other hand, motivated by the success of quantum tomographic techniques in optics \cite{jmo},
more direct methods for measuring the quantum state of a BEC were recently proposed
\cite{eurobec,tomobecal,eric}.

In optical tomography, the key point is the use of a reference field, 
namely the local oscillator \cite{jmo}.
The latter, prepared in a coherent state, allows one to probe the desired state through 
the measurement of a set of probabilities
\cite{ray}.
However, a difficulty arises if we try to adapt the same technique to a BEC. 
In fact, while in optics it is easy to
obtain a coherent reference field (e.g. from a laser), the same is not actually 
available for atoms. 
Nevertheless, recent progresses in this direction seem promising \cite{boser,dunn}. 
Hence, in the present paper, we provide a detailed study
of the possibility to reconstruct the quantum state of a BEC which 
includes both the scenarios: when a reference field is available and when it is not.

The paper is organized as follows: in Section II we review the tomographic principle and
consider a suitable operator transform on the atomic system. In Section III, we consider the
case of state reconstruction in the absence of a reference field, while in Section IV the
opposite case is analyzed.
Finally, in Section V we comment the numerical results, and in Section VI we conclude
with a brief discussion.

\section{The basic model}

Recently, we established ~\cite{ourjmo} a quite general 
principle
of constructing measurable probabilities, which 
determine completely the quantum state
in the tomographic approach. For a  more refined treatment
see Ref.~\cite{dar}. 

Let us consider a quantum state described by the density
operator ${\hat\rho}$, which is a nonnegative 
Hermitian operator, i.e., 
\begin{equation} \label{rhoher}
{\hat\rho}^{\dag}={\hat\rho}, \qquad \mbox {Tr}
\,{\hat\rho}=1\,,
\end{equation}
and 
\begin{equation} \label{rhovv}
\langle v\mid{\hat\rho}\mid v \rangle=\rho_{v,v}\geq 0.
\end{equation}
We label the vector basis $\mid v \rangle $ in the 
space of pure quantum states
by the index $v$ which may represent any degrees of 
freedom of the system
under consideration.  
Formula~(\ref{rhovv}) can be 
rewritten by using the
Hermitian projection operator 
\begin{equation}  \label{Piv}
{\hat \Pi}_{v}=\mid v \rangle\langle v \mid,
\end{equation}
in the following form 
\begin{equation}  \label{rhovvPi}
\rho_{v,v}=\mbox{Tr}
\left({\hat \Pi}_{v}{\hat\rho}\right)\,.
\end{equation}

On the other hand, in the space of states, there will be a family of unitary 
transformation operators 
${\hat U}(\sigma)$ depending on some parameters 
$\sigma=(\sigma_1,\ldots,\sigma_k\ldots)$, 
that can be sometimes identified with a 
group-representation operators. 
It was shown \cite{ourjmo} that known tomography schemes 
can be
considered from the viewpoint of group theory 
by using appropriate groups. 
More recently this concept has been developed 
obtaining an elegant group theoretical 
approach to quantum state measurement \cite{dar}.
Here, we formulate the tomographic approach in the 
following way. Let us
introduce a ``transformed density operator''
\begin{equation}  \label{rhosig}
{\hat\rho}_{\sigma}={\hat U}^{-1}(\sigma){\hat\rho} 
{\hat U}(\sigma).
\end{equation}
Its diagonal elements are still nonnegative probabilities 
\begin{equation}  \label{wz}
\langle v \mid{\hat\rho}_{\sigma} \mid v \rangle
\equiv w(v,\sigma)\ge 0\,.
\end{equation}
These probabilities are functions of stochastic variable(s) $v$ and parameter(s) $\sigma$.
As a consequence of the unit trace of the density 
operator, they also fulfill
the normalization condition 
\begin{equation}  \label{normz}
\int \,dv\,w\left(v,\sigma\right)=1.
\end{equation}
Of course, in the case of discrete indices, the integral 
in Eq.~(\ref{normz}) 
should be replaced by a sum over discrete variables.

The l.h.s. of Eq.~(\ref{wz}) can be interpreted as the probability 
density for the measurement of the observable ${\hat V}$
(the operator whose eigenstates are given by $|v\rangle$) 
in an ensemble of
transformed reference frames labeled by the index $\sigma$, 
if the state
${\hat\rho}$ is given. Along with this interpretation, 
one can also
consider the transformed projector
\begin{equation}\label{Pizsig}
{\hat \Pi}_v(\sigma)
={\hat U}(\sigma){\hat\Pi}_v {\hat U}^{-1}(\sigma)\,,
\end{equation}
in terms of which the expression~(\ref{wz}) for the
probability $w \left(v,\sigma\right)$ takes the form 
\begin{equation}\label{wzPi}
w\left(v,\sigma\right)=
\mbox{Tr}\left[ {\hat\rho}\,{\hat\Pi}_v(\sigma) \right]\,.  
\end{equation}
These probability densities are also called 
``marginal" distributions 
as a generalization of the concept introduced by 
Wigner~\cite{wig}.
The tomography schemes are 
based on the possibility to find the inverse of 
Eq.~(\ref{wzPi}). If it is
possible to solve Eq.~(\ref{wzPi}), considering 
the probability 
$w\left(v,\sigma \right)$ as a known function and 
the density matrix as an unknown
operator, the quantum state can be reconstructed in terms
of measurable positive definite probability distributions.
This is 
the essence of state reconstruction 
techniques.

Specifically, we consider two atomic sources whose atoms
(described by two bosonic modes ${\hat b}_1$ and ${\hat b}_2$) can be mixed through an 
atomic beam splitter \cite{bs}, 
and assume that successively a phase shift $\phi$ can also be introduced between them.
We shall specify these modes later.
At the output, a detection of the number of atoms in both modes can be performed.
This amounts to the possibility of measuring the probability distributions related to
the transformed state
\begin{equation}\label{prob}
{\hat\rho} \to {\hat U}(\theta,\phi) \, {\hat\rho} \, {\hat U}^{\dag}(\theta,\phi) \,,
\end{equation}
where the transformation operator is given by
\begin{equation}\label{KTh}
{\hat U}(\theta,\phi)=\exp\left\{
-i\frac{\theta}{2}\left[b_1^{\dag}b_2 \exp(-i\phi)
+b_1b_2^{\dag}\exp(i\phi)\right]\right\}\,.
\end{equation}
Here, $\cos^2(\theta/2)$ represents the transmission coefficient at the beam splitter.
Eq.~(\ref{prob}) plays the same role of Eq.~(\ref{rhosig}), and,
in the spirit of the tomographic principle, the set of ``marginals" 
associated to the transformed state will allow us to recover the original state.
In the next two Sections, as anticipated in the Introduction, 
we shall distinguish two situations.

\section{Case I}

We first treat the case where a reference field is not available.
All we can do in this case is to consider two condensates belonging to the two modes
${\hat b}_1$, ${\hat b}_2$, 
and put the constraint of total particle number conservation, i.e. $[{\hat\rho},{\hat N}]=0$,
in order to infer their (joint) state.
The latter is assumed to be a generic two-mode state of the type
\begin{equation}\label{twoms}
|\Psi \rangle=\sum^N_{n=0} c_n \, |N-n\rangle_1\,|n\rangle_2\,.
\end{equation}
At this stage we use the formal equivalence between the algebra for two 
harmonic oscillators and that for angular momentum \cite{groups}.
We write the state $|n\rangle_1\,|N-m\rangle_2=|j+m\rangle_1\,|j-m\rangle_2$
as a spin state $|m\rangle$, where $j=N/2$ and $m=n-j$ 
($m=-j,-j+1,\ldots,j-1,j$).
The $j+1$ states $|m\rangle$ have all the properties of the eigenstates
of ${\hat J}^2$ and ${\hat J}_z$ where 
\begin{equation} \label{angular}
{\hat J}_+={\hat J}^{\dag}_-={\hat b}_1^{\dag}{\hat b}_2\,,\quad
{\hat J}_z=\frac{1}{2}\left(
{\hat b}_1^{\dag}{\hat b}_1-{\hat b}_2^{\dag}{\hat b}_2
\right)\,,\quad
{\hat J}^2={\hat J}_z^{2}+\frac{1}{2}\left({\hat J}_+{\hat J}_-
+{\hat J}_-{\hat J}_+\right)
\,.
\end{equation}
The effect of the beam splitter, including the phase shift,
is a rotation by an angle $\theta$ about an axis
${\bf u}_{\phi}={\bf u}_x \cos\phi-{\bf u}_y \sin\phi$
of the angular momenta ${\hat{\bf J}}$, i.e.
\begin{equation}\label{Jrot}
{\hat U}(\theta,\phi)=\exp\left(-i\theta{\hat{\bf J}}\cdot {\bf u}_{\phi}\right)\,.
\end{equation}
On the other hand, the rotation (\ref{Jrot}) can be specified by means of the Wigner-D function
\cite{groups}
\begin{equation}\label{UequivD}
\langle m'| {\hat U}(\theta,\phi) |m\rangle \equiv 
{\cal D}^{(j)}_{m'\,m}(\psi=0,\theta,\phi) \,,
\end{equation}
where now
$\psi$, $\theta$, $\phi$  represent the Euler's angles.
Then, the probability of $j+m$ counts at the first detector and $j-m$ 
at the second one is given by
\begin{equation}\label{wsex}
w\left(m,\theta,\phi\right) = 
\sum^j_{m_1=-j}\,\sum^j_{m_2=-j}
\,{\cal D}_{m\,m_1}^{(j)}(\psi,\theta,\phi)\,
\rho^{(j)}_{m_1\,m_2}\,{\cal D}_{m\,m_2}^{(j)\,*}
(\psi,\theta,\phi)\,.
\end{equation}
The measurement of the atomic number in both modes guarantees a unit efficiency.
In fact, data for which the sum of counts is not $N$ can be disregarded.
Moreover, in Eq.~(\ref{wsex}) we have left the argument $\psi$ unspecified
in the r.h.s. and omitted it in the l.h.s. since
${\cal D}^{(j)}_{m\,m'}\propto \exp(-i m \psi)$:
the marginal distribution only depends on the two angles 
$\theta$ and $\phi$.

Following~\cite{dodman,olga} we will derive 
the expression for the
density matrix of a spin state in terms of measurable 
probability
distributions. 
This can be done by using the known integral product of three Wigner-D functions over the 
rotation group and the orthogonality of the Wigner-3$j$ symbols
${\cal W}^{j_1\,j_2\,j_3}_{m_1\,m_2\,m_3}$ \cite{groups}.
Finally, the density matrix elements can be expressed in terms of the marginal 
distribution as
\begin{eqnarray}\label{rhows}
\rho_{m_1\,m_2}^{(j)}&=&  
(-1)^{m_2}\sum_{j'=0}^{2j}\,\sum_{m'=-j'}^{j'}\,(2j'+1)^2
\sum_{m=-j}^{j} \int (-1)^{m} w\left(m,\theta,\phi\right)\, 
\nonumber\\
&\times& 
{\cal D}_{0\,m'}^{(j')}(\psi,\theta,\phi)\;
{\cal W}^{j\,j\,j'}_{m\,-m\,0}\;{\cal W}^{j\,j\,j'}_{m_1\,-m_2\,m'}
\,\frac{d\Omega}{8\pi^2}
\end{eqnarray}
where the integration is performed over 
the rotation parameters, i.e.
\begin{equation}\label{solid}
\int\,d\Omega\,=\int_0^{2\pi}\,d\psi\,
\int_0^{\pi}\,\sin\theta\,d\theta\,\int_0^{2\pi}\,d\phi\,.
\end{equation}
Thus, Eq.~(\ref{rhows}) can be used to sample two-mode BEC density matrix elements starting 
from the measurable probabilities $w(m,\theta,\phi)$ and some known functions.

\section{Case II}

Recent progress in the generation of an atomic coherent source \cite{boser} makes us hope about
the possibility to have an atomic reference field \cite{dunn}.
Thus, we shall consider in this section the first mode as the condensate to be investigated, 
and the second one coming from a coherent atomic source.

Let ${\hat\rho}$ be the state of mode 1 we want to reconstruct
and ${\overline\beta}$ the coherent state characterizing mode 2. Then, the probability
of counting $n$ atoms in mode 1 for $\theta=\pi/2$, 
will be
\begin{equation}\label{Pnbdef}
w(n,\beta)={\rm Tr}\left[{\hat U}^{-1}(\theta=\pi/2,\phi)\,
{\hat\rho}\,|{\overline\beta}\rangle_2{}_2\langle{\overline\beta}|\, 
{\hat U}(\theta=\pi/2,\phi)\,
|n\rangle_1{}_1\langle n|
\right]
\,,
\end{equation}
where $\beta=|{\overline\beta}|\exp(i\varphi)$, $\varphi=\arg{\overline\beta}-\phi+\pi/2$.
This corresponds to the probability distribution for 
the measurement of the displaced number operator
${\hat D}^{\dag}(\beta) 
{\hat b}^{\dag}_1{\hat b}_1 {\hat D}(\beta)$
analogously to the Photon
Number Tomography \cite{PNT}. In that case, however, one has to collect number distributions 
by spanning the whole complex plane $\beta$; here, instead, we will simplify the procedure
(see also \cite{opat}).

Of course, the number of atoms in the condensate, though not fixed, will be finite,
thus, it happens that
$\langle k|\rho|m\rangle=0$ for
$k,m>N_1$, with
$N_1$  a suitable estimation of the maximum number of the condensed atoms. 
By virtue of this
assumption we can rewrite Eq. (\ref{Pnbdef}) as 
\begin{eqnarray}
w(n,\beta) & = & \exp(-|\beta|^2)n!\sum_{k,m=0}^{N_1}\langle k|\rho|m\rangle
\frac{1}{\sqrt{k!m!}}
|\beta|^{m+k-2n}\exp[i(m-k)\varphi]
L^{(m-n)}_n\left(|\beta|^2\right)
\nonumber \\
 & & \times L^{(k-n)}_n\left(|\beta|^2\right)\,,
\label{pnal}
\end{eqnarray}
where $L^{(m)}_n$ are the associated Laguerre polynomials.

Let us now consider, for a given value of  $|\beta|$, the function $w(n,\beta)$ as function of
$\varphi$ and calculate the coefficients of the Fourier expansion
\begin{equation}\label{pns}
w^{(s)}(n,|\beta|)=\frac{1}{2\pi}\int_0^{2\pi}
d\varphi\,w(n,\beta)\exp(is\varphi)\,,
\end{equation}
$(s=0,1,2,\ldots)$. By combining Eqs.~(\ref{pnal}) and (\ref{pns}), we get
\begin{equation}\label{pnsrho}
w^{(s)}(n,|\beta|)=\sum_{m=0}^{N_1-s}{\cal A}^{(s)}_{n,m}(|\beta|)
\langle m+s|\rho|m\rangle\,,
\end{equation}
where
\begin{equation}\label{Gs}
{\cal A}^{(s)}_{n,m}(|\beta|)=\exp(-|\beta|^2)n!
\frac{1}{\sqrt{(m+s)!m!}}
|\beta|^{2(m-n)+s}
L^{(m-n)}_n\left(|\beta|^2\right)
L^{(m+s-n)}_n\left(|\beta|^2\right)\,.
\end{equation}

If the distribution $w(n,\beta)$ is measured for
$n=0,1,\ldots,N$  ($N\ge N_1$), then Eq. (\ref{pnsrho})
represents for each value of $s$ a system of $(N+1)$ linear equations
between the $(N+1)$ measured quantities and the $(N_1+1-s)$
unknown density matrix elements.
Therefore, to obtain the latter we only need to invert the system \cite{Robi}
\begin{equation}\label{rhobec}
\langle m+s|\rho|m\rangle=\sum_{n=0}^{N}{\cal M}^{(s)}_{m,n}(|\beta|)w^{(s)}(n,|\beta|)\,,
\end{equation}
where the matrices ${\cal M}$ are given by ${\cal M}=({\cal A}^T{\cal A})^{-1}{\cal A}^T$. It is
possible to see that such matrices satisfy the relation
\begin{equation}\label{FGmatr}
\sum_{n=0}^{N}{\cal M}^{(s)}_{m',n}(|\beta|){\cal A}^{(s)}_{n,m}(|\beta|)=\delta_{m,m'}\,,
\end{equation}
for $m,m'=0,1,\ldots,N_1-s$, which means that from the exact probabilities satisfying Eq.
(\ref{pnsrho}) the correct density matrix is obtained.
By combining Eqs. (\ref{rhobec}) and (\ref{pns}) we find that
\begin{equation}\label{samplebec}
\langle m+s|\rho|m\rangle=\frac{1}{2\pi}\sum_{n=0}^{N}\int\,d\varphi\,
{\cal M}^{(s)}_{m,n}(|\beta|)\exp(is\varphi)w(n,\beta)\,,
\end{equation}
which may be regarded as the formula for the direct sampling of the condensate density matrix.
In particular we see that the determination of the state of the condensate only requires the value
of $\varphi$ (i.e. the phase between reference and condensate field) to be varied.
Moreover, the presented reconstruction procedure involves Laguerre polynomials in place of
additional summations,which guarantee a better stability in the numerical manipulation of large set
of data, with respect to analogous methods~\cite{opat}.

Finally, the non unit efficiency $\eta$ in the detection process can be accounted for
by considering a binomial convolution of the ideal probability \cite{scully}
\begin{equation} \label{weta}
w_{\eta}(k,\beta)=\sum_{n=k}^{\infty}
\left(
\begin{array}{c}
n\\k
\end{array}
\right)
\eta^k (1-\eta)^{n-k} w(n,\beta)\,,
\end{equation}
and the consequent modification of the matrix ${\cal A}$.

\section{Numerical Results}

It is plausible, and it has been already suggested \cite{squee,eric}, 
that the state of a condensate with
repulsive collisions be a squeezed state with reduced number fluctuations.
Hence in the following we will consider this situation.
The single 1-mode state can be written as 
\begin{equation}\label{psisq}
|\Psi\rangle=\sum_{n=0}^{\infty} c_n |n \rangle\,,
\end{equation}
where the coefficients $c_n$ are given by \cite{schleich}
\begin{equation} \label{cn}
c_n=\left(\frac{2}{r+1}\right)^{1/2}r^{1/4}\left(\frac{r-1}{r+1}\right)^{n/2}
\left(2^n n!\right)^{-1/2} H_n\left(\sqrt{\frac{2 
r^2}{r^2-1}}\,x_0\right)
\exp\left(-\frac{r}{r+1}\,x_0^2\right)\,,
\end{equation}
where $r$ is the squeezing parameter, $x_0$ is the (real) displacement and 
$H_n$ denotes the Hermite polynomials.

A phase space representation of this state can be given by the 
$Q$-function~\cite{wfun}
\begin{equation}\label{Qdef}
Q(\alpha)=\langle\alpha|{\hat\rho}|\alpha\rangle\,,
\end{equation}
where $\alpha$ is the complex amplitude of a coherent state.
It yields
\begin{equation}\label{Qone}
Q(\alpha)
=\exp(-|\alpha|^2)\left|
\sum_{n=0}^{\infty}\frac{(\alpha^*)^n}{\sqrt{n!}} \, c_n
\right|^2\,.
\end{equation}
For the case discussed in Sec. III, we have to 
consider~\cite{squee,eric} a two-mode squeezed state written in the 
angular momentum representation, {\it i.e.}
\begin{equation}\label{psisq12}
|\Psi\rangle={\cal N} \sum_{m=-j}^{j} c_{j+m} |m \rangle\,,
\end{equation}
where $\cal N$ is a normalization factor and the coefficients 
$c_{j+m}$ are given in Eq.~(\ref{cn}).
Since the quantity $x_0^2+(r^2-1)/4r$ represents the mean number
of atoms in mode 1, it must
be smaller then the total number of atoms $N$.
Furthermore, the atomic coherent state basis for a system of angular momentum
$j$ is defined by~\cite{arecchi}
\begin{eqnarray}
|\theta,\phi\rangle & = &
\sum_{m=-j}^{j}{\cal D}^{(j)}_{m,-j}(\psi=0,\theta,\phi) |m\rangle
\nonumber \\
 & = & \sum_{m=-j}^{j}
\left(
\begin{array}{c}
2j\\
m+j
\end{array}
\right)^{1/2}
\left(\sin\frac{\theta}{2}\right)^{j+m}
\left(\cos\frac{\theta}{2}\right)^{j-m}
\exp(-im\phi) |m\rangle\,,
\label{coh12}
\end{eqnarray}
then, one can define the $Q$-quasiprobability 
distribution analogously to Eq.~(\ref{Qdef})
\begin{equation} \label{Qdef12}
Q(\theta,\phi)=
\langle \theta,\phi |{\hat\rho}| \theta, \phi\rangle\,.
\end{equation}
For the state considered in Eq. (\ref{psisq12}), it becomes
\begin{equation} \label{Q12}
Q(\theta,\phi)=\left|
\sum_{m=-j}^{j}{\cal D}^{(j)\,*}_{m,-j}(\psi=0,\theta,\phi) c_{j+m}
\right|^2\,.
\end{equation}
This is shown in Fig.~1(a). Instead, Fig.~1(b) displays the $Q$-function 
calculated from the 
reconstructed density matrix elements.
We may see that the method of Sec. III is quite accurate, apart 
from some background noise. 

Analogously, in Fig.~2(a) we have plotted the ideal Wigner 
function~\cite{wfun} of Eq.~(\ref{Qone}),
while Fig.~2(b) is its reconstructed version.
In this case, the statistical error depends on the chosen value of $|\beta|$.
For $|\beta|$ close to zero the diagonal density matrix elements can be determined 
very precisely, whereas the off-diagonal elements strongly fluctuate.
The opposite happens by increasing $|\beta|$.
To compensate for the fluctuations, the number of measurement events must be increased.
Another source of error stems from the truncation of the reconstructed density matrix at
the value $N_1$.

It is worth comparing the previous results with those obtained in the 
case of a number state. For this purpose, we show in Fig.~3(a) the 
ideal Wigner function for a number state. In this case, the Wigner 
function becomes negative, displaying the highly non classical 
character of a Fock state. As above, in Fig.~3(b) we show the Wigner 
function obtained by applying the reconstruction method of Sec.~IV to 
such a state: the two pictures are practically indistinguishable, 
showing the accuracy of the present method.

Finally, as an instructive comparison we show in Fig.~4 the same 
Wigner function of Fig.~2(a) calculated when the state reconstruction takes place
with a random phase relation between probe and condensate. 
As can be seen the state becomes randomized and diffused in phase, but 
its Wigner function remains positive. This figure should be contrasted
with Fig.~3. In this case the apparent $U(1)$ symmetry is not pertaining to the state \cite{for}, 
but rather due to the measurement method, which implies 
a preparation of the (same) state at each experimental run.

\section{Conclusion}

To conclude, we have studied, through numerical simulations, the possibilities of a tomographic
approach to the quantum state of a Bose-Einstein condensate. 
We have considered two possible scenarios, whether an atomic reference field is available or not.
The corresponding methods turn out to be accurate and robust to detection 
inefficiency, and allow one to distinguish among various possible 
quantum states of the condensate.

It is worth noting that the studied techniques allow
direct sampling of the density matrix elements avoiding 
any ambiguities in the reconstruction procedure 
due to singularities \cite{dalp}.

The key point remains the 
possibility to have a reference field and/or its state preparation. 
Furthermore, we note the necessity to
deal with a relatively small number of atoms in order to implement 
efficiently the numerical algorithms.
In spite of these difficulties, 
we retain the possibility of measuring the true density matrix 
of a condensate accessible
and worth considering. 

Finally, we would like to remark that the presented procedures could 
also be considered for other
fields like high energy heavy ion collisions where pions can condense as well
\cite{pion}.

\section*{Acknowledgments}

The authors gratefully acknowledge discussions with R. Onofrio.
This work has been partially supported by the Istituto Nazionale di Fisica della Materia 
under the Advanced Research Project ``Cat", the Ministero 
dell'Universit\`a e della Ricerca Scientifica e Tecnologica, and the 
European Union.

\begin{figure}
%\centerline{\epsfig{figure=fig1a.eps,width=7cm} \ \ 
%\epsfig{figure=fig1b.eps,width=7cm}}
\caption{A squeezed 2-mode state. (a) Ideal $Q$-function when the 
displacement parameter is $x_{0}=\protect\sqrt{5}$ and the squeezing 
parameter is $r=e$. (b) The corresponding $Q$-function reconstructed through 
the method of Sec.~III. To obtain this figure we have simulated experimental
data by adding to each probability $w$ a noise term with a Gaussian distribution,
the latter having a width proportional to the ratio between the probability itself
and the number of runs for given parameters.}
\label{fig1}
\end{figure} 

\begin{figure}
%\centerline{\epsfig{figure=fig2a.eps,width=7cm} \ \ 
%\epsfig{figure=fig2b.eps,width=7cm}}
\caption{A squeezed state for the single mode. (a) Ideal Wigner function when the 
displacement parameter is $x_{0}=\protect\sqrt{3}$ and the squeezing 
parameter is $r=e$. (b) The corresponding Wigner function reconstructed through 
the method of Sec.~IV. The reconstruction parameters are 
$|\beta|=1.1$, $\eta=0.9$, and $3\times 10^{5}$ simulated experimental 
data per each phase have been used (see text).}
\label{fig2}
\end{figure} 

\begin{figure}
%\centerline{\epsfig{figure=fig3a.eps,width=7cm} \ \ 
%\epsfig{figure=fig3b.eps,width=7cm}}
\caption{A number state for the single mode. (a) Ideal Wigner function 
for the Fock state $|n\rangle=|5\rangle$. (b) The corresponding Wigner 
function reconstructed through the method of Sec.~IV. The reconstruction
parameters are $|\beta|=0.3$, $\eta=0.9$, and $3\times 10^{5}$ simulated
experimental data per each phase have been used (see text).}
\label{fig3}
\end{figure} 

\begin{figure}
%\centerline{\epsfig{figure=fig4.eps,width=7cm}}
\caption{A squeezed state with a random phase between $0$ and $2\pi$. 
Here, a value of the displacement parameter $x_{0}=\protect\sqrt{5}$
has been used.}
\label{fig4}
\end{figure}

\end{document}